\documentclass[twocolumn,nofootinbib]{revtex4}
\usepackage{amssymb}
\newcommand*\de{\mathrm{d}}
\newcommand*\De{\mathrm{D}}
\renewcommand*\epsilon{\varepsilon}
\renewcommand*\phi{\varphi}
\renewcommand*\theta{\vartheta}
\begin{document}

\title{On the teleparallel limit of Poincar\'e gauge theory} 
\author{ M. Leclerc}
\date{11 January 2005}
\affiliation{Section of Astrophysics and Astronomy, Department of Physics,
University of Athens, Greece}
\begin{abstract}
We will address the question of the consistency of teleparallel  
theories in presence of spinning matter which has been a 
controversial subject 
of discussion over the last twenty years. 
We argue that the origin of the problem is not simply  the 
symmetry or asymmetry of the stress-energy tensor of the matter fields, 
which has been recently analyzed by several authors, but arises at a more 
fundamental level, namely from the invariance of the field equations under
a frame change, a problem that has been discussed long time ago by 
Kopczynski in the framework of the teleparallel equivalent of general 
relativity.      
More importantly, we show that the problem is not only confined to the purely  
teleparallel theory  but arises actually in every Poincar\'e gauge theory 
that admits 
a teleparallel geometry in the absence of spinning sources, i.e. in its 
classical limit.

\end{abstract}
\maketitle
\section{Introduction}

Recently \cite{1,2}  there has been a revival of the 
discussion on whether or not 
the Dirac field can be consistently coupled to gravity in the 
framework of the teleparallel equivalent of general relativity (TEGR). 
The authors of \cite{2} came to the conclusion that the theory, with 
the usual minimal coupling prescription (which we consider exclusively 
in this paper), is not consistent. The reason for this is simply the 
fact that the theory leads to a symmetric Einstein equation and 
thus requires the right hand side of this equation, namely the 
stress-energy tensor of the Dirac particle, to be symmetric too. 
Clearly, the stress-energy tensor of the Dirac particle, as well as 
of any other particle with intrinsic spin (when minimally coupled), 
is not symmetric by itself. In other words, requiring its symmetry 
is a constraint on the fermion field. Especially, the spin tensor would have  
to be conserved (covariantly), a condition that is not even satisfied 
in the absence of gravitational fields. 

On the other hand, the inconsistency of  TEGR 
has already been claimed twenty years ago in \cite{3} (see also \cite{4}-
\cite{7}),
using a different argumentation. It has been noted that  TEGR 
Lagrangian possesses a symmetry that is not inherit of the matter 
Lagrangian of a spinning particle. Namely, the Lagrangian and the 
field equations (in the absence of spinning matter) are invariant under 
what is called a frame 
transformation, i.e. a Lorentz transformation of the tetrad field 
with the connection held fixed (see equation (7) below). As a consequence 
of this symmetry, the torsion tensor  is not entirely determined by the field 
equations. Since spinning matter fields do not present the same 
invariance (in other words, they couple directly to the torsion), their 
behavior, when treated as test fields, cannot be predicted by the theory. 
Actually, the authors of \cite{3}-\cite{7} do not confine their 
analysis to  TEGR. Rather, they consider the so-called 
one-parameter teleparallel Lagrangian, which leads to the most general
teleparallel geometry that is consistent with the experimental situation. In
this article, we  confine ourselves to those Lagrangians that present a
classical limit that is completely equivalent to general relativity. The
discussion is easily generalized
\newpage  
\setlength{\parindent}{0mm} to the more general case (see remark at the 
end of section III).  

\setlength{\parindent}{3mm}
The scope of this article is to show that the problem described in 
\cite{1} and \cite{2} is actually directly related to the frame 
invariance of the teleparallel Lagrangian analyzed in \cite{3} and 
that it is not confined to the teleparallel equivalent of general relativity, 
but is present in 
any Poincar\'e gauge theory that leads to a  teleparallel geometry 
(with equations equivalent to those of general relativity)  
in its classical limit, i.e. in the absence of spinning matter fields. 

In order to fix our notations and conventions, we briefly review 
the basic concepts of Riemann-Cartan geometry which is the 
basis of Poincar\'e gauge theory. For a detailed introduction, consult  
 the standard reference \cite{8}. 
Latin letters from 
the beginning of the alphabet ($a,b,c\dots $) run from 0 to 3 and 
are (flat) tangent space indices. Especially, $\eta_{ab}$ is the 
Minkowski metric $diag(1,-1,-1,-1)$ in tangent space. Latin letters 
from the middle of the alphabet ($i,j,k \dots $) are indices in a curved 
spacetime with metric $g_{ik}$. We introduce the independent gauge 
fields, the tetrad  $e^a_m $ and the connection 
$\Gamma^{ab}_{\ \ m}$ (antisymmetric in $ab$) and 
the correspondent field 
strengths, the curvature and torsion tensors
\begin{eqnarray}
R^{ab}_{\ \ lm} &=& \Gamma^{ab}_{\ \ m,l} - \Gamma^{ab}_{\ \ l,m} 
                     + \Gamma^a_{\ cl}\Gamma^{cb}_{\ \ m}   
- \Gamma^a_{\ cm}\Gamma^{cb}_{\ \ l}  \\ 
T^a_{\ lm} &=& e^a_{m,l} - e^a_{l,m} + e^b_m \Gamma^a_{\ bl}
- e^b_l \Gamma^a_{\ bm}. 
\end{eqnarray}
The spacetime connection $\Gamma^i_{lm} $ and the spacetime metric $g_{ik}$
can now be defined through 
\begin{equation}
e^a_{m,l} + \Gamma^a_{\ bl} e^b_m = e^a_i \Gamma^i_{ml} \ \ \text{and} \ \ 
e^a_ie^b_k \eta_{ab} = g_{ik}.  
\end{equation}
It is understood that there exists an inverse to the tetrad, such that 
$e^a_i e_b^i = \delta^a_b $. It can now be shown that the connection 
splits in two parts, 
\begin{equation}
\Gamma^{ab}_{\ \ m} = \hat \Gamma^{ab}_{\ \ m} +
K^{ab}_{\ \ m},
\end{equation} 
such that $\hat \Gamma^{ab}_{\ \ m}$ is torsion free and the contortion 
$K^a_{\ bm}$ is related to the torsion through 
$T^a_{\ ik} = K^a_{\ bi}e^b_k- K^a_{\ bk}e^b_i$. Especially, 
the spacetime connection $\hat \Gamma^i_{lm}$ constructed from 
$e^a_{m,l} + \hat \Gamma^a_{\ bl} e^b_m = e^a_i \hat \Gamma^i_{ml}$
is just the Christoffel connection of general relativity, a function of 
the metric only. 

All quantities constructed with the torsion free connection 
$\hat \Gamma^{ab}_{\ \  m}$ or $\hat \Gamma^i_{lm}$ will be denoted with 
a hat. Thus, for instance,  $\hat R^i_{\ lkm}$ is 
the usual Riemann curvature tensor. 

The gauge fields $e^a_m $ and $\Gamma^{ab}_{\ \ m} $ are 
vector fields with respect to the spacetime index $m$. Under a local 
gauge transformation in tangent space, $\Lambda^a_{\ b}(x^m)$, 
they transform as
\begin{equation}
e^a_m \rightarrow \Lambda^a_{\ b}e^b_m, \ \ \ \Gamma^a_{\ bm} \rightarrow 
\Lambda^a_{\ c}\Lambda^d_{\ b} \Gamma^c_{\ dm} - 
\Lambda^a_{\ c,m} \Lambda^{\ c}_{b}. 
\end{equation}

The transformation (5) is the basis of Poincar\'e gauge theories. 
Under this transformation, the torsion and the curvature transform 
homogeneously. We will refer to it as Poincar\'e gauge transformation, 
although it 
is actually only the Lorentz part of a Poincar\'e transformation after 
having fixed the translational part to the so called physical gauge. 
This conception of the Poincar\'e transformation is described in 
\cite{9}. (For a fundamental treatment in a more general framework, 
see \cite{10}.)  Every Lagrangian, 
gravitational or not, should be invariant under (5). 

In addition, one can consider the pure Lorentz gauge transformations
\begin{equation}
e^a_m \rightarrow e^a_m, \ \ \ \Gamma^a_{\ bm} \rightarrow 
\Lambda^a_{\ c}\Lambda^d_{\ b} \Gamma^c_{\ dm} - 
\Lambda^a_{\ c,m} \Lambda^{\ c}_{b},  
\end{equation}
as well as the frame transformations
\begin{equation}
e^a_m \rightarrow \Lambda^a_{\ b}e^b_m, \ \ \ \Gamma^a_{\ bm} \rightarrow 
\Gamma^a_{\ bm}.  
\end{equation}
Clearly, neither (6) nor (7)  are symmetries of the Dirac Lagrangian 
(always speaking of the minimally coupled Lagrangian) 
nor of the Einstein-Cartan Lagrangian for instance. Note also that the 
transformation (5), (6) and (7) are not independent. Clearly, a Lorentz transformation 
 (6) followed by a frame transformation (7) (with the same parameters) 
is equivalent to a Poincar\'e transformation (5).

In the next section, we will investigate under which conditions 
the stress-energy tensor of the matter fields is symmetric. Then, in 
section III, we construct the family of Lagrangians that present a 
teleparallel limit in the spinless case and discuss  
the problem of the inconsistency of such theories in the presence of spinning 
particles in relation with  their 
invariance under a frame change (7).

\section{Frame invariance and symmetry of the stress-energy tensor}

We now deduce the conservation laws that follow from the symmetries 
(5), (6) and (7) of  a general matter Lagrangian density $\mathcal L_m$, 
which may depend on $e^a_m, \Gamma^{ab}_m$ (as well as on their 
derivatives) and on matter 
fields that we summarize under the notation $\psi$. 

As usual, we use the canonical definitions of the stress-energy tensor 
and of the spin density under the form 
\begin{displaymath}
T^a_{\ m} = \frac{1}{2e} \frac{\delta \mathcal L_m}{\delta e^m_a}, \ \  
\sigma_{ab}^{\ \ m} = \frac{1}{e} \frac{\delta \mathcal L_m}{\delta 
\Gamma^{ab}_{\ \ m}}.
\end{displaymath}

We consider infinitesimal transformations $\Lambda^a_{\ b} = \delta^a_b 
+ \epsilon^a_{\ b}$ with $\epsilon^{ab} = - \epsilon^{ba}$. (As tangent 
space indices, $a,b\dots$ are  and lowered with $\eta_{ab}$.) 

The Poincar\'e transformation (5) now takes the form 
\begin{equation}
\delta \Gamma^{ab}_{\ \ m} = - \epsilon^{ab}_{\ \ ,m} + \epsilon^a_{\ c}
\Gamma^{cb}_{\ \ m} + \epsilon^b_{\ c}\Gamma^{ac}_{\ \ m}, \ \  
\ \delta e^a_m = \epsilon^a_{\ c} e^c_m. 
\end{equation}
The inverse of the tetrad transforms  with the inverse 
transformation, i.e. $\delta e^m_a = \epsilon_a^{\ c}e_c^m$. 
The matter action $S_m = \int \mathcal L_m \de^4 x$ 
therefore undergoes the following change (up to a boundary term):  
\begin{eqnarray*}
\delta S_m &=& \int (\frac{\delta \mathcal L_m}{\delta e^m_a}\delta e^m_a 
+ \frac{\delta \mathcal L_m }{\delta \Gamma^{ab}_{\ \ m} }\delta
\Gamma^{ab}_{\ \ m} ) \de^4 x  \\
&=& \int e (2 T^{[ab]} + D_m \sigma^{abm}) \epsilon_{ab} \de^4 x,  
\end{eqnarray*}
where $D_m $ is the covariant derivative that acts with $\Gamma^{ab}_{\ \ m}$ 
on the tangent space indices and with $\hat \Gamma^i_{kl} $ (torsion less) 
on the spacetime indices. 
We conclude that, if the matter Lagrangian 
possesses the symmetry (5), we have the following (well known) conservation 
law
\begin{equation}
D_m \sigma^{abm} + 2 T^{[ab]} = 0. 
\end{equation}
If the matter fields $\psi$ too are subject to a 
gauge transformation   
(for instance $\delta \psi = i\epsilon^{ab} \sigma_{ab}\psi $ in the 
Dirac case, with the Lorentz generators $\sigma_{ab}$), the action 
undergoes 
an additional change $\frac{\delta \mathcal L_m}{\delta \psi}\delta \psi$, 
but this does not contribute, due to the  field equations of the matter
fields,  
which are derived from  $\frac{\delta \mathcal L_m}{\delta \psi}= 0 $. 

Clearly, the same argument if applied to the transformation (6) instead of 
(5) leads to 
$D_m \sigma^{abm} = 0$ and if applied to the frame 
change (7) to $T^{[ab]} = 0$. Since we consider only Lagrangians that 
possess the Poincar\'e symmetry, 
the symmetry (7) will imply the symmetry (6) and vice versa. Therefore, 
we can state that if the Lagrangian is frame invariant, then we 
have the conservation laws 
\begin{equation}
D_m \sigma^{abm} = 0 \ \ \  \text{and} \ \ \    T^{[ab]} = 0. 
\end{equation}

Until now, we have considered only the matter part of the Lagrangian. 
Similar arguments can be applied to the gravitational 
Lagrangian $\mathcal L_0$ itself, which depends only 
on $e^a_m, \Gamma^{ab}_{\ \ m}$
and their first derivatives. If we define $C_{ab}^{\ \ m} = - e^{-1} 
\delta \mathcal 
L_0 / \delta \Gamma^{ab}_{\ \ m} $ and $E^a_m = -(2e)^{-1} \delta 
\mathcal L_0 / \delta e^m_a$,  the gravitational field equations 
arising from $\mathcal L = \mathcal L_0 + \mathcal L_m$ have the 
form 
\begin{equation}
E^a_{\ m} = T^a_{\ m}, \ \ \ C_{ab}^{\ \ m} = \sigma_{ab}^{\ \ m}, 
\end{equation}
where as usual we refer to the first equation as Einstein equation 
and to the second one as Cartan equation. 

Using the same argumentation as before, we can show that every 
Poincar\'e invariant Lagrangian $\mathcal L_o$ will satisfy 
the Bianchi identity  
\begin{equation}
D_m C^{abm} + 2 E^{[ab]} = 0. 
\end{equation}
If $\mathcal L_0$ is in addition frame invariant, we have the relations 
\begin{equation}
D_m C^{abm} = 0 \ \ \ \text{and}\ \ \   E^{[ab]} = 0. 
\end{equation}

\section{Poincar\'e gauge theory with teleparallel limit}

A major problem in Poincar\'e gauge theory consists in reducing  
the 11 parameter Lagrangian (see \cite{11} for instance) 
to those Lagrangians which are compatible with the classical 
experimental situation. Since our experiments until today 
are confined to the metrical structure of spacetime, we can be sure 
to be in agreement with the experiments if the metric obeys the 
classical Einstein equations $\hat G_{ik} = T_{ik}$. Therefore, we will 
look for Lagrangians whose Einstein equation $E^a_{\ m}= T^a_{\ m}$, 
in the case of a vanishing spin density 
of the matter fields, reduces to $\hat G_{ik} = T_{ik}$. We know 
at least two such theories, namely general relativity (GR) itself 
(which can be seen as the classical limit of Einstein-Cartan (EC) theory, 
the zero spin condition leading to zero torsion) and the teleparallel 
equivalent of GR (TEGR) where $R^{ab}_{\ \ lm} = 0$. 

One goal of Poincar\'e gauge theory  is to generalize the 
above theories to allow for both  
dynamical torsion and curvature. This means that we have to include 
 at least one term quadratic in the curvature into the Lagrangian. If we seek 
for a classical limit with zero torsion, this term will certainly contribute
to the Einstein equation even in the classical limit, except if it is 
of a very special (and unnatural) 
form like $R^{i[klm]}R_{i[klm]}$ or $R^{[lm]}R_{[lm]}$ 
(here, $[ikl]$ means total anti-symmetrization of the three indices). Such  
terms actually depend only on torsion derivatives and vanish in the 
zero torsion limit via the Bianchi identities in Riemannian space. 

On the other hand, if we are looking for a teleparallel limit in the zero spin 
case, we can add all kinds  of terms quadratic in the curvature, 
$R_{ik}R^{ik},\ R^2 \dots$, 
without changing 
the classical limit of the theory. Such terms  will lead only 
to contributions that 
vanish in the zero curvature limit. These are the Lagrangians we 
 investigate in this paper. 

Apart from the quadratic curvature terms, we have to modify 
the TEGR Lagrangian such that it is suitable for a first order 
variation without the use of Lagrange multipliers (see \cite{12}). 
The  suitable  Lagrangian can be found (in a more general framework) 
in \cite{13}. It consists of  
the sum of the teleparallel and the EC Lagrangian ($eL_0 =  \mathcal L_0$), 
\begin{equation}
 L_0 =   R - \frac{1}{4}T^{ikl}T_{ikl}-\frac{1}{2}T^{ikl}T_{lki} 
+ \frac{1}{2}T^k_{\ ik}T^{mi}_{\ \ m}.   
\end{equation}
This Lagrangian, apart from a divergence term, 
is essentially the Einstein-Hilbert Lagrangian (expressed in 
terms of the tetrad) (see \cite{13} or \cite{14}).     
It leads, in the absence of spinning matter, to the 
GR equation 
$\hat G_{ik} = T_{ik}$ and the Cartan equation is identically fulfilled. 
(In other words, $\delta \mathcal L_0 / \delta \Gamma^{ab}_{\ \ m} = 0$.)
This means that $\Gamma^{ab}_{\ \ m} $ remains completely undetermined. 

Note that (14) is frame invariant and consistently, the Einstein tensor 
is symmetric. Let us now look at the Lagrangian 
\begin{equation}
L = L_0 + a R^{ab}_{\ \ lm}R_{ab}^{\ \ lm} + L_m,
\end{equation}
with $L_0$ from (14) and $L_m$ some matter Lagrangian. The field equations 
now read 
\begin{eqnarray}
\hat G_{ik} &=& \tau_{ik} + T_{ik},\\ 
D_m R^{ablm} &=& \sigma^{abl}, 
\end{eqnarray}
with $\tau_{ik} = -2a[R^{ab}_{\ \ li}R_{ab\ k}^{\ \ l}- (1/4)R^{ab}_{\ \ lm}
R_{ab}^{\ \ lm}]$. We chose (15) as an illustrative example because of its 
simple structure. Its field equations are exactly those of an
Einstein-Yang-Mills system. Instead of $R^{ab}_{\ \ ik}R_{ab}^{\ \ ik}$ 
we can take any combination of quadratic curvature terms, because in the 
following, we are interested mainly in the classical, teleparallel limit. 

Clearly, if the source is spinless, we get $R^{ab}_{\ \ lm} = 0$ as 
ground state solution. With this solution, we have $\tau_{ik} = 0$, 
and (16) reduces to the Einstein equation of GR. 

We now come to the discussion of references \cite{1} and \cite{2}. The 
main statement in \cite{2} is the fact that  TEGR  is 
not consistent when coupled to the Dirac particle because its Einstein 
equation  has a 
symmetric left hand side but the stress-energy tensor of the Dirac particle 
is asymmetric. We agree completely with this view, but we will show that 
the roots of the problem can be traced back to the frame invariance  
not only of the field equations, but of their classical limit (i.e. even 
in the absence of the Dirac particle as source). Therefore, the 
discussion should not be confined to the symmetry properties of $T_{ik}$. 

Indeed, the Lagrangian (15) is again frame invariant, and thus equation 
(16) has the same symmetry problem as the corresponding one 
considered in \cite{1} and 
\cite{2}. However, this problem can be cured very easily: We simply add 
a term $b R^2$ (with the curvature scalar $R = e^i_ae^k_bR^{ab}_{\ \ ik}$) 
to (15). This term is clearly not frame invariant 
(although Poincar\'e invariant) and thus breaks the unwanted symmetry. 
(Any other quadratic curvature term that is not frame invariant does the 
same job. Again, the term $R^2$ serves as illustrative example.) Especially, 
we will get an additional asymmetric contribution 
$\sim R(4R_{ik} - g_{ik}R) $ to (16), allowing therefore for 
an asymmetric $T_{ik}$. Further, we get a contribution to the Cartan equation
(17) of the form $\sim \De_i (e^i_{[a} e^k_{b]} R)$. Therefore, from the point 
of view of the discussion in \cite{1} and \cite{2}, which focuses on 
the symmetry properties of the Einstein equation, the problem has been solved. 

However, in the absence of  spinning sources, we get as before the  
ground state 
solution $R^{ab}_{\ \ ik} = 0$, and therefore the (teleparallel) 
Einstein equation $\hat G_{ik} = T_{ik}$. 
Note that $T_{ik}$ is now supposed to be symmetric, since the source is 
classical. These equations are  once again frame invariant. 

What does that mean? Well, let us fix the Poincar\'e gauge by 
imposing $\Gamma^{ab}_{\ \ m} = 0$. Then, from the Einstein equation, 
we can determine the metric $g_{ik}$. But the tetrad field will be 
determined only up to a Lorentz transformation 
$e^a_m \rightarrow \Lambda^a_{\  b}e^b_m$. This is the problem that has 
been discussed in \cite{3} twenty years ago in the framework of 
the teleparallel equivalent of general relativity. For classical matter, 
this is not a problem, because it couples to the 
metric alone. Especially, the geodesics of a classical test particle 
will not depend on the gauge choice. 
However, spinning particles couple directly to the tetrad (or to the torsion,
which is not a tensor under (7)) 
and the (semiclassical) trajectory of a test particle entering our fields, 
as well as its  spin precession 
equation, will depend on the specific frame we choose. 
We can therefore not take the point of view that all the solutions that 
differ only by a frame change are equivalent. 

We can even reduce the whole discussion to the complete groundstate 
of the field equations. The groundstate solution of the 
Einstein equation is $g_{ik} = \eta_{ik}$
and that of the Cartan equation is $R^{ab}_{\ \ lm} = 0$. Without 
physical consequences, we can fix the Poincar\'e gauge 
by the requirement $\Gamma^{ab}_{\ \ m} = 0$. Obviously, this state 
is invariant under (7). We can therefore determine neither the 
tetrad, nor the torsion (which is not a tensor under (7)). 
These fields however are measurable since they couple to spinning 
particles. Clearly, this problem arises in any theory whose 
field equations reduce in the classical limit to $R^{ab}_{\ \ lm} = 0$ 
and $\hat G_{ik} = T_{ik}$.

Finally, there is the possibility of adding the term 
$\lambda T^{[ikl]}T_{[ikl]}$ (the square of the totally antisymmetric 
torsion part) to the Lagrangian. This changes the classical limit 
slightly but in a way consistent with the experimental situation, for 
an arbitrary constant $\lambda$. (The so-called one-parameter teleparallel
theory, see \cite{15}). This 
breaks the frame invariance of  
the classical 
limit (and even of the groundstate), 
but it has been shown in \cite{3} that there is a remaining 
invariance of the form $e^a_m \rightarrow \Lambda^a_{\ b}e^b_m$ with 
$\Lambda^a_{\ b}$ a special  Lorentz transformation that leaves the 
axial torsion part unaffected. Therefore, taking into account this 
new term  would solve the problem for the Dirac 
test particle, which couples only to the axial torsion, but if we 
consider higher spin fields or macroscopic spin polarized bodies, the 
problem reappears, since the latter   
couple also to the other torsion parts (vector and tensor)
which remain undetermined  (see 
\cite{16} for semi-classical equations of momentum propagation and of 
precession for general spinning test bodies). The complete discussion, in the 
framework of the purely teleparallel theory, can 
be found in \cite{3}. The results of \cite{3} have been confirmed and analyzed 
  in greater detail in the follow-up articles \cite{4}-\cite{7}. 
In order to solve the problem completely, the torsion has to be fixed 
(determined) completely even in the classical limit (and especially in 
the ground state of the theory). Therefore, if we want a Poincar\'e theory 
to have a general relativity limit in the spinless case, this limit cannot
 correspond to a teleparallel geometry, but should be described by a fixed 
torsion, most probably  $T^a_{\ ik} = 0$, i.e. a Riemannian geometry.  

\section{Conclusion}

As a result, we conclude that  the 
teleparallel equivalent of general relativity 
is not consistent in presence of minimally coupled 
spinning matter. We showed that the argument given in \cite{2}, i.e. 
that the Einstein equation has a symmetric left hand side whereas the 
stress-energy tensor of spinning matter field is not symmetric, 
actually has its roots in the frame invariance of the teleparallel Lagrangian 
discussed in \cite{3}. 

Furthermore, we could show that every Poincar\'e gauge theory that 
leads, in the absence of spinning matter,  
to a teleparallel geometry with an Einstein equation equivalent to GR 
suffers from the same inconsistency. Even if the Lagrangian itself 
is not frame invariant, the field equations in their classical limit 
will be frame invariant again. A spinning test particle entering 
these fields however will couple directly to the torsion (which is not 
a tensor under the frame change), and its behavior (spin precession, 
trajectory\dots) will depend on the arbitrary choice of a specific frame.  

The problem with such theories has also been analyzed in \cite{17}, 
based on a  completely different  
argumentation (3+1-decomposition). The conclusions are 
similar, however, our argumentation is much simpler 
and shows clearly  which class 
of theories suffers from the inconsistency and why there is a relation 
to the symmetry of the Einstein equation.

\end{document}